\UseRawInputEncoding
\documentclass[10pt, superscriptaddress, twocolumn, amsmath, amssymb, aps,pra, notitlepage,longbibliography]{revtex4-1} 
\usepackage{graphicx,graphics,epsfig,subfigure,times,bm,bbm,amssymb,amsmath,amsfonts,amsthm,mathrsfs,MnSymbol,physics}
\usepackage[all]{xy}
\usepackage[normalem]{ulem}
\usepackage{slashed}
\usepackage{dcolumn}
\usepackage{tabularx}
\usepackage{amsopn}
\usepackage{color}
\usepackage[dvipsnames,svgnames,table]{xcolor}
\usepackage[english]{babel}
\usepackage{verbatim}
\definecolor{darkblue}{rgb}{0.0,0.0,0.3}
\usepackage[colorlinks=true,
            linkcolor=red,
            urlcolor= darkblue,
            citecolor=blue]{hyperref}

\begin{document}

\title{Non-Hermitian engineering of superfluidity in a Rashba spin-orbit-coupled Fermi gas}
\author{Pingcheng Zhu}
\affiliation{Institute for Quantum Science and Technology, Shanghai University, Shanghai, 200444, China}
\author{Lihong Zhou}
\email{lihongzh@shu.edu.cn}
\affiliation{Institute for Quantum Science and Technology, Shanghai University, Shanghai, 200444, China}
\author{Jianxin Zhong}
\email{jxzhong@shu.edu.cn}
\affiliation{Institute for Quantum Science and Technology, Shanghai University, Shanghai, 200444, China}
\begin{abstract}
We investigate superfluid pairing in a two-dimensional Rashba spin-orbit-coupled Fermi gas subject to spin-selective one-body loss. Within the non-Hermitian mean-field framework, we self-consistently solve the gap and number equations and find that moderate dissipation can significantly enhance the pairing
gap, resulting in a pronounced nonmonotonic dependence on the dissipation strength. Dissipation also provides an additional control parameter for driving the system across the BCS-BEC crossover. We further analyze the
quasi-particle spectrum and identify two distinct superfluid regimes characterized by one and three exceptional rings, separated by an exceptional spectral transition. Interestingly,   dissipation can  enhance both pairing channels  while simultaneously inducing a momentum-dependent phase twist in the triplet component. These results demonstrate that
spin-selective dissipation provides a versatile non-Hermitian control knob for manipulating superfluid pairing, spectral structure, and crossover physics in spin-orbit-coupled quantum gases.
\end{abstract}

\date{\today}
\maketitle

\section{Introduction}
The study of open quantum systems has driven a paradigm shift in modern physics, moving the focus from isolated Hermitian settings to non-Hermitian (NH) platforms where dissipation and gain are no longer viewed solely as sources of decoherence but can act as active control knobs~\cite{Ashida2020, Ganainy2018}. A central frontier lies in extending NH physics to strongly correlated many-body physics, particularly fermionic superfluids with dissipation interactions or particle losses with unconventional phase transitions~\cite{Ueda2019, Iskin2021,Shi2024,koga2024,Wang2025,zhu2026}, and the emergence of non-diagonalizable exceptional points, lines, or surfaces in the quasiparticle spectrum~\cite{Okugawa2019, Kozii2024}. This framework has been widely extended to low-dimensional lattices and chiral systems, uncovering a rich spectrum of phenomena, including robust edge modes under the non-Hermitian skin effect (NHSE)~\cite{wang2018,Lee2019,Gon2020,Wang2026} and non-Hermitian p-wave pairing~\cite{ p2020,Ji2025}. Within this broader NH many-body framework, a particularly fertile direction is the interplay between NH potentials and spin-orbit coupling (SOC), which can give rise to even richer physics.

The experimental realization of synthetic SOC has revolutionized the study of ultracold Fermi gases~\cite{Gong2011, Jiang2011, Ren2022}. In particular, Rashba SOC breaks inversion symmetry, lifts spin degeneracy, and mixes spin-singlet and spin-triplet pairing~\cite{Hu2011,Zhai2011,He2012,Shi2016}. It also enhances the pairing gap and superfluid transition temperature even in the weakly interacting BCS regime ~\cite{Zhai2011, Hu2011} and fundamentally reshapes the BCS-BEC crossover dynamics~\cite{Gong2011,He2012, Shi2016,Iskin2026}.

When SOC systems are coupled to an environment, dissipation can give rise to qualitatively new non-Hermitian phenomena. Recent studies have demonstrated that engineered dissipation, such as an effective imaginary magnetic field generated by spin-selective atom loss, can enhance fermion  superfluidity~\cite{Zhou2019,Zhao2023}. Moreover, introducing dissipation into Raman lattices can produce non-Hermitian SOC, which not only facilitates the formation of  bound molecules over a broader parameter regime~\cite{Zhou2020} but also induces the  NHSE, whereby a macroscopic number of eigenstates accumulate near the system boundaries~\cite{,nat2025,Li2023,Zhou2022,Li2024}. These developments highlight that dissipation is not merely a decoherence mechanism but can serve as a versatile resource for engineering exotic quantum states and topological phenomena. Despite substantial progress in understanding non-Hermitian $s$-wave and $p$-wave pairing, the interplay between non-Hermiticity and synthetic gauge fields--especially in spin-orbit-coupled Fermi systems under dissipation remains largely unexplored. In particular, it remains unclear how non-Hermiticity modifies pairing symmetry, triplet correlations, and BCS-BEC crossover physics in spin-orbit-coupled Fermi gases.

 Motivated by these recent advances, this paper investigates a two-dimensional Rashba spin-orbit coupling Fermi gas subject to one-body dissipation.  By extending the mean-field BCS-BEC crossover theory into the non-Hermitian regime, we derive the self-consistent conditional mean-field solution, and find that moderate dissipation significantly enhances the pairing gap, with a pronounced non-monotonic dependence on the loss strength. Moreover, dissipation can serve as a new tuning knob to drive the system across the BCS-BEC crossover. We further analyze the quasiparticle spectrum and identify two distinct exceptional superfluid regimes, one featuring a single exceptional ring and the other three exceptional rings separated by an exceptional spectral transition. We also show that dissipation simultaneously enhances both singlet and equal-spin triplet pairing channels and generates a momentum-dependent phase twist in the triplet component. These results establish spin-selective dissipation as a versatile non-Hermitian control knob for manipulating superfluid pairing, spectral structure, and crossover physics in spin-orbit-coupled quantum gases.

The rest of the paper is organized as follows. In Sec.~\ref{model}, we describe the model and theoretical methods. Sec.~\ref{selfconsist} presents the self-consistent numerical results. Sec.~\ref{pairing} discusses the pairing behavior, and we conclude with a discussion and conclusion in Sec.~\ref{conclusion}.

\section{Model Hamiltonian}\label{model}
The system under consideration is a two-dimensional uniform s-wave interacting Fermi gas, subject to both Rashba SOC in the $x-y$ plane and a spin-dependent atom loss described by the non-Hermitian term 
 $i\Gamma\sigma_z$. According to the post-selecting approximation, the effective Hamiltonian of the system is given by $\hat{\mathcal{H}} = \hat{\mathcal{H}}_0+\hat{\mathcal{H}}_{\mathrm{int}}$, where the single particle Hamiltonian $\hat{\mathcal{H}}_0=\sum_{\textbf{k}ss'}c^\dagger_{\textbf{k}s}[\xi_\textbf{k}I+\alpha(k_y\sigma_x-k_x\sigma_y)+i\Gamma\sigma_z]_{ss'}c_{\mathbf{k}s'}$, ${s=\uparrow,\downarrow}$ are the pseudospin of the atoms, $\alpha$ and $\Gamma$ are the SOC and dissipation strength, $I$ and $\sigma_{i}$ are the unit and Pauli matrix, respectively. $\xi_\textbf{k}=\hbar^2{k}^2/2m-\mu$, $\mu$ is the chemical potential. The $s$-wave interaction term $\mathcal{H}_\text{int}=(g/S)\sum_{\textbf{k}\textbf{k}'}c^\dagger_{\textbf{k}\uparrow}c^\dagger_{-\textbf{k}\downarrow}c_{-\textbf{k}'\downarrow}c_{\textbf{k}',\uparrow}$ , with  $\frac{1}{g}=-\frac{1}{S}\sum_{\bf{k}}\frac{1}{2\epsilon_{\bf{k}}+E_b}$~\cite{He2012}, where $E_b$ is the two-body binding energy without SOC case and  $S$ is the quantization area. 

We first study the single-particle physics in the absence of  Fermi sea with  $\mu=0$, then the single particle spectrum is
\begin{align}
\epsilon_{\pm}(\textbf{k}) =\frac{\hbar^2{k}^2}{2m}\pm\sqrt{\alpha^2 k^2-\Gamma^2}
\end{align}
where $k = \sqrt{k_x^2 + k_y^2}$. The energy minimum $\epsilon_{\text{min}}=-\frac{m\alpha^2}{2\hbar^2}+\frac{\hbar^2\Gamma^2}{2m\alpha^2}<0$ for $\Gamma<\Gamma_c=\frac{m\alpha^2}{\hbar^2}$. The single-particle Hamiltonian possesses a global generalized
parity-time ($\mathcal{PT}$) symmetry defined by
$\mathcal{PT}=\sigma_x\mathcal{K}$, where $\sigma_x$ exchanges the two
pseudospin states and $\mathcal{K}$ denotes complex conjugation. Although the global Hamiltonian $\mathcal{H}_0$ respects $\mathcal{PT}$ symmetry, its eigenstates display spontaneous $\mathcal{PT}$ symmetry breaking that is momentum-dependent. Bounded by the exceptional ring at $k_{EP} = \Gamma/\alpha$, where both eigenvalues and eigenvectors coalesce, the single-particle energy spectrum is divided in momentum space into two distinct regions: one preserving $\mathcal{PT}$ symmetry ($k>k_{EP}$) and one breaking it ($k<k_{EP}$).

In the following, we study the many-body physics. Based on the non-Hermitian mean-field approximation~\cite{Takemori2024}, the $s$-wave superfluid pair potentials are $\Delta=(g/S)\sum_{\textbf{k}}{}_{L}\langle c_{-\textbf{k}\downarrow}c_{\textbf{k}\uparrow}\rangle_{R}$, $\tilde\Delta=(g/S)\sum_{\textbf{k}}{}_{L}\langle c^\dagger_{\textbf{k}\uparrow}c^\dagger_{-\textbf{k}\downarrow}\rangle_{R}$ and interaction term turns to be $\mathcal{H}_\text{int}=\sum_\textbf{k}(\tilde{\Delta}c_{-\textbf{k}\downarrow}c_{\textbf{k}\uparrow}+\Delta c^\dagger_{\textbf{k}\uparrow}c^\dagger_{-\textbf{k}\downarrow})-\Delta\tilde{\Delta}S/g$. Under the Nambu spinor basis $\Psi_\textbf{k}=(c_{\textbf{k}\uparrow},c_{\textbf{k}\downarrow},c^\dagger_{-\textbf{k}\uparrow},c^\dagger_{-\textbf{k}\downarrow})^{\text{T}}$, the Hamiltonian is $\mathcal{H}=\frac{1}{2}\sum_{\textbf{k}}\Psi^\dagger_\textbf{k}M_\textbf{k}\Psi_\textbf{k}+\sum_\textbf{k}\xi_\textbf{k}-\Delta\tilde\Delta S/g$, where matrix
\begin{align}
M_\textbf{k}&=
    \begin{pmatrix}
        \mathcal{H}_0(\textbf{k})&i\sigma_y\Delta \\
        -i\sigma_y\tilde\Delta&-\mathcal{H}^{\text{T}}_0(-\textbf{k})
    \end{pmatrix}
    \label{eq2}
\end{align}
preserves the particle-hole symmetry. The quasiparticle excitation energies are $E^\lambda_{\textbf{k}\pm}=\lambda\sqrt{\xi_\textbf{k}^2+\alpha^2k^2-\Gamma^2+\Delta \tilde{\Delta}\pm2\mathcal{B}_\textbf{k}}$,
where $\lambda = \pm$ correspond to the particle and hole branches, $\mathcal{B}_\textbf{k} = \sqrt{\alpha^2 k^2 \xi_{\mathbf{k}}^2-\Gamma^2(\xi_{\mathbf{k}}^2 + \Delta \tilde{\Delta}) }$. For $\alpha = 0$ and $\Gamma = 0$, it reduces to $E_{\mathbf{k}}^{\lambda} = \lambda \sqrt{\xi_{\mathbf{k}}^2 + \Delta\tilde{\Delta}}$, recovering the standard BCS result. For the present model, the two saddle-point equations admit a symmetric solution with \(|\Delta|=|\tilde\Delta|\). A global \(U(1)\) gauge transformation can then be used to fix their common phase, yielding $\Delta=\tilde\Delta\equiv\Delta_0$. In the following calculation, we take Fermi energy $E_F$ and Fermi wave vector $k_F$ as the units of energy and momentum, which is related to the total particle density $n$ via $k_F^2=2\pi n$.

\begin{figure}[t] 
 \centering
\includegraphics[width=1\columnwidth]{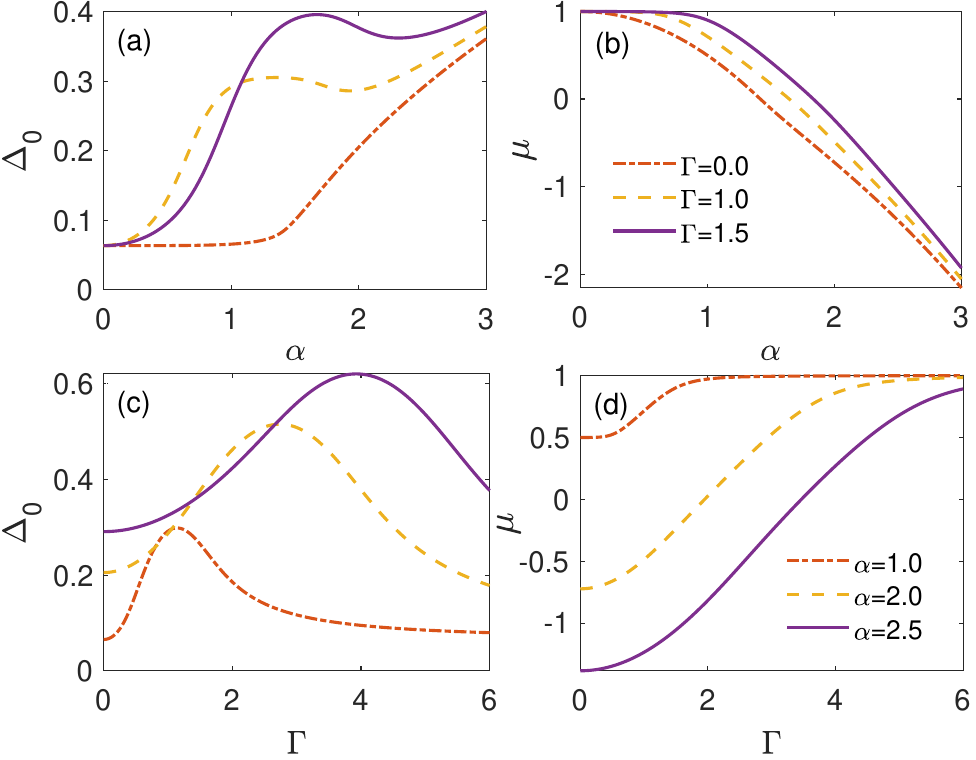} 
\caption{Plots of (a) order parameter $\Delta_0$ and (b) chemical potential $\mu$ with respect to SOC strength $\alpha$ for different dissipation $\Gamma=(0,1,1.5)E_F$.  Plots of (c) $\Delta_0$ and (d) $\mu$ with respect to $\Gamma$ for different $\alpha =(1,2,2.5)E_F/k_F$. Here the interaction strength $E_b=0.002 E_F$.}
\protect\label{fig:Fig1Main}
\end{figure}
\section{Self-consistent Solution}\label{selfconsist}
 
\subsection{Ground state}
By evaluating the  expectation values of order parameter and total particle number
$N=
\sum_{\mathbf{k},\sigma}
{}_{L}\langle c^\dagger_{\textbf{k}\sigma}c_{\textbf{k}\sigma}\rangle_{R}$. Using the biorthogonal
Bogoliubov transformation~\cite{Brody2014} and  two-dimensional renormalization, we then obtain the gap equation
\begin{equation}
\sum_\textbf{k}\frac{1}{2\epsilon_\textbf{k}+E_b} = \sum_{\textbf{k},\eta=\pm} \left(1 - \frac{\eta \Gamma^2}{\mathcal{B}_\textbf{k}}\right) \frac{1}{4E^+_{\textbf{k}\eta}}
\end{equation}
and the number equation
\begin{equation}
N =  \sum_{\textbf{k},\eta=\pm} \left[\frac{1}{2}-\left( \frac{\eta (\alpha^2 k^2 - \Gamma^2)}{\mathcal{B}_\textbf{k}} + 1 \right)  \frac{\xi_\textbf{k}}{2E^+_{\textbf{k}\eta}}\right]
\end{equation}
 
Solving these equations self-consistently determines the chemical potential $\mu$ and order parameter $\Delta_0$ in the ground state. 
We begin with the weakly interacting regime on the BCS side, where the two-body binding energy is chosen as $E_b=0.002E_F$;  results for the strongly interacting regime are presented in the appendix ~\ref{a:2}. Fig. \ref{fig:Fig1Main}(a) and (b) show the dependence of the pairing gap $\Delta_0$ and chemical potential $\mu$ on the Rashba SOC strength $\alpha$ for several dissipation strengths $\Gamma$. In the absence of dissipation, there is a characteristic value roughly at $\alpha k_F/E_F\sim1.3$, below which the change of $\Delta_0$ with $\alpha$ is small, while above this value the increasing of $\Delta_0$ becomes significant. This behavior is consistent with previous studies of three dimensional Fermi gas~\cite{Gong2011,Zhai2011}. In contrast, a finite dissipation strength significantly enhances the pairing gap, and this enhancement becomes more pronounced with increasing SOC, contrary to the Hermitian case with real Zeeman field~\cite{Gong2011}. 
In Fig.~\ref{fig:Fig1Main}(b) we plot $\mu$ as a function of $\alpha$, with increasing $\alpha$, $\mu$ decreases continuously and eventually becomes deep negative, indicating the formation of tightly bound molecular pairs or BCS-BEC crossover. Not surprisingly, these behaviors are not qualitatively different from the case without dissipation.
\begin{figure}[t] 
 \centering
\includegraphics[width=1\columnwidth]{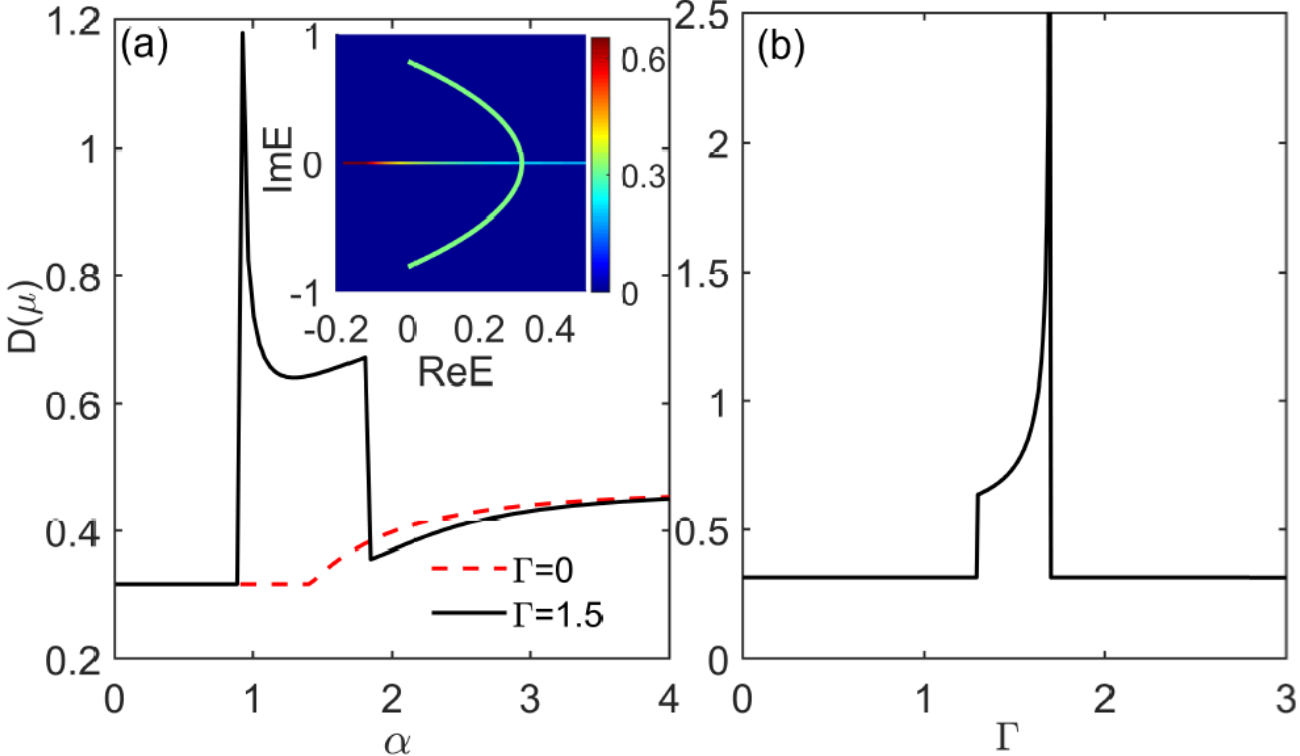} 
\caption{Plot of density of states (DOS) at Fermi energy $\mu$ as functions of (a) SOC strength $\alpha$ with  different dissipation $\Gamma=0E_F$(dashed line), $\Gamma=1.5E_F$(solid line), the inset is the DOS in the complex energy plane for $\alpha =1E_F/k_F$, $\Gamma=0.8E_F$, and (b)$\Gamma$ with fixed $\alpha=1E_F/k_F$. All plots with parameter $E_b=0.002E_F$.}
\protect\label{fig:DOS}
\end{figure}
Next we examine the effect of dissipation $\Gamma$ at fixed $\alpha$. As shown in Fig.\ref{fig:Fig1Main}(c), the pairing gap $\Delta_0$ exhibits a non-monotonic dependence on $\Gamma$. Starting from the Hermitian limit, increasing $\Gamma$ initially enhances $\Delta_0$ and drives it to a maximum. Further increasing $\Gamma$ gradually suppresses this enhancement, and $\Delta_0$ approaches the value in the absence of both SOC and dissipation in the strong-loss limit, corresponding to the quantum Zeno effect. This non-monotonic behavior reflects the cooperative and competing effects of SOC and dissipation on the quasiparticle spectrum and pairing structure: at moderate dissipation, the non-Hermitian spectral
reconstruction and SOC-induced spin mixing act cooperatively to favor pairing, whereas sufficiently strong loss competes with and
weakens the SOC-induced spin mixing that underlies the pairing enhancement. Additionally, the corresponding evolution of the chemical potential $\mu$ in Fig.\ref{fig:Fig1Main}(d) shows that, for sufficiently strong SOC, increasing dissipation can drive the system from a  BEC-like regime back toward a  BCS-like regime. Therefore, non-Hermiticity introduces a new dimension for manipulating pairing regimes beyond conventional tuning parameters such as interaction strength and SOC.
\subsection{Single-particle density of states}
To further understand the microscopic origin of the dissipation-induced modification of superfluidity, we investigate the single particle density of states (DOS) in the complex energy plane~\cite{Takemori2025}
\begin{align}
D(E) = \frac{1}{S}  \sum_{{\bf{k}},\lambda=\pm} \delta(\text{Re}E-\text{Re}\epsilon_\lambda({\bf{k}}))\delta(\text{Im}E-\text{Im}\epsilon_\lambda({\bf{k}}))
\end{align}
where $\delta(x)$ is the delta function. Within the exceptional ring, the single-particle energies develop finite imaginary components, corresponding to gain and loss branches, while outside the ring the spectrum remains real. Consequently, varying the SOC strength or dissipation modifies the relative weight of the non-Hermitian region in momentum space and leads to a controllable reconstruction of the effective DOS.

Fig. \ref{fig:DOS}(a) and \ref{fig:DOS}(b) present the DOS at the Fermi energy as functions of $\alpha$ and $\Gamma$, respectively. In the absence of dissipation, increasing $\alpha$ modifies the DOS through the conventional Rashba band reconstruction. When finite dissipation is introduced, the DOS is further enhanced owing to the emergence of complex-energy states. This enhancement correlates with the increase in the pairing gap observed in Fig.~\ref{fig:Fig1Main}, suggesting that the non-Hermitian spectral reconstruction provides an important microscopic mechanism for dissipation-enhanced superfluidity.
The complex energy distribution shown in the inset of Fig.~\ref{fig:DOS} demonstrates that the single-particle states are continuously redistributed around the exceptional ring boundary. This feature highlights the unique role of exceptional structures in interacting non-Hermitian quantum systems.
\begin{figure}[h] 
 \centering
\includegraphics[width=0.85\columnwidth]{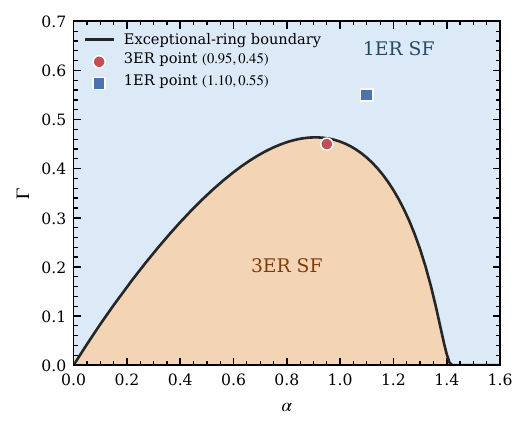} 
\caption{Self-consistent exceptional-ring superfluid phase diagram in the$(\alpha$--$\Gamma)$ plane at $E_b=0.002E_F$.}
\protect\label{fig:ESFdiagram}
\end{figure}
\subsection{Exceptional-ring superfluid regimes}
\label{subsec:ER_phases}
At finite dissipation strength, the Bogoliubov quasiparticle branches can develop exceptional degeneracies in momentum space~\cite{Takemori2025}. For a fixed particle-hole index $\lambda$, the two  branches $E_{\mathbf{k}+}^{\lambda}$ and $E_{\mathbf{k}-}^{\lambda}$ coalesce when the discriminant $\mathcal{B}_{\mathbf{k}}$ vanishes, yielding $\alpha^{2}k^{2}\xi_{\mathbf{k}}^{2}
=\Gamma^{2}\left(\xi_{\mathbf{k}}^{2}+\Delta_{0}^{2}\right)
$. Owing to the rotational symmetry of the system, each positive real solution defines an exceptional ring in the two-dimensional momentum plane.
\begin{figure}[ht] 
 \centering
\includegraphics[width=0.85\columnwidth]{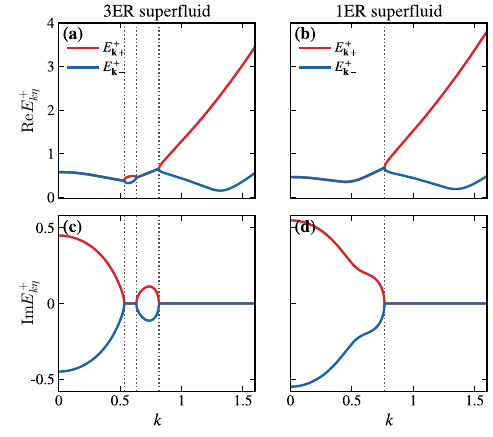} 
\caption{Real and imaginary parts of the Bogoliubov quasiparticle energies $E_{\mathbf{k}\pm}^{+}$. (a) Three-exceptional-ring regime with $\alpha=0.95E_F/k_F$, $\Gamma=0.45E_F$. (b) One-exceptional-ring regime with $\alpha=1.10E_F/k_F$, $\Gamma=0.55E_F$. The interaction strength is fixed at $E_b=0.002E_F$.}
\protect\label{fig:ESF}
\end{figure}
In the dimensionless units adopted below, where $\xi_{\mathbf{k}}=k^{2}-\mu$, we obtain
$
\left(k^{2}-\mu\right)^{2}
\left(\alpha^{2}k^{2}-\Gamma^{2}\right)
-\Gamma^{2}\Delta_{0}^{2}=0.$
Introducing $x=k^{2}$, the radial exceptional condition reduces to the cubic equation
\begin{equation}
\mathcal{P}(x)
\equiv
(x-\mu)^{2}\left(\alpha^{2}x-\Gamma^{2}\right)
-\Gamma^{2}\Delta_{0}^{2}
=0.
\label{eq:ER_cubic}
\end{equation}
We now investigate how the self-consistent pairing field reconstructs the exceptional structure of the Bogoliubov spectrum. As shown in Fig.~\ref{fig:ESFdiagram}, for each point in the $(\alpha,\Gamma)$ parameter plane, we substitute the solutions of $\Delta_{0}$ and $\mu$ into Eq.~(\ref{eq:ER_cubic}). Since $\mathcal{P}(x)$ is cubic in $x=k^{2}$, the non-Hermitian superfluid can exhibit either one or three concentric exceptional rings, which we refer to as the one exceptional ring superfluid (1ER SF) and the three exceptional ring superfluid (3ER SF), respectively. In the normal state limit $\Delta_{0}\rightarrow0$, Eq.~(\ref{eq:ER_cubic}) contains the single-particle exceptional-ring condition $\alpha k=\Gamma$, whereas a finite pairing field reconstructs this ring through particle-hole hybridization and allows additional exceptional rings to emerge. In the 1ER SF regime, the associated exceptional ring separates two regions with distinct complex spectral structures. In the 3ER SF regime, two additional positive roots emerge, producing three concentric exceptional rings. This richer structure originates from the interplay between the non-monotonic radial dispersion $\xi_{\mathbf{k}}=k^{2}-\mu$, particle-hole hybridization induced by $\Delta_{0}$, and the dissipation term $i\Gamma$. In particular, the pairing term $-\Gamma^{2}\Delta_{0}^{2}$ in Eq.~(\ref{eq:ER_cubic}) shifts the extrema of $\mathcal{P}(x)$ and enables the creation of an additional pair of exceptional rings.

The boundary between the 1ER SF and 3ER SF regimes is determined by the coalescence of two radial solutions. It therefore satisfies the double-root conditions
\begin{equation}
\mathcal{P}(x_{c})=0,
\qquad
\left.\frac{\partial\mathcal{P}(x)}
{\partial x}\right|_{x=x_{c}}=0,
\label{eq:ER_boundary}
\end{equation}
where $k_{c}=\sqrt{x_{c}}$ is the critical ring radius.  Acrossing this boundary, a pair of exceptional rings is created or annihilated, while the remaining ring evolves continuously. As shown in Figs.~\ref{fig:ESF}, the real and imaginary part of quasiparticle branches reconnect differently in the 1ER and 3ER regimes, and the branch coalescence points coincide with the exceptional ring radii given by Eq.~(\ref{eq:ER_cubic}).The change in the number of exceptional rings reflects a reconstruction of the complex Bogoliubov spectrum rather than a conventional symmetry-breaking transition. In particular, although the topology of the exceptional set changes, the self-consistent order parameter and chemical potential may remain continuous across the boundary. We therefore characterize the 1ER-3ER boundary as an exceptional spectral transition.

\section{Pairing correlations}\label{pairing}
\begin{figure}[t] 
 \centering
\includegraphics[width=0.85\columnwidth]{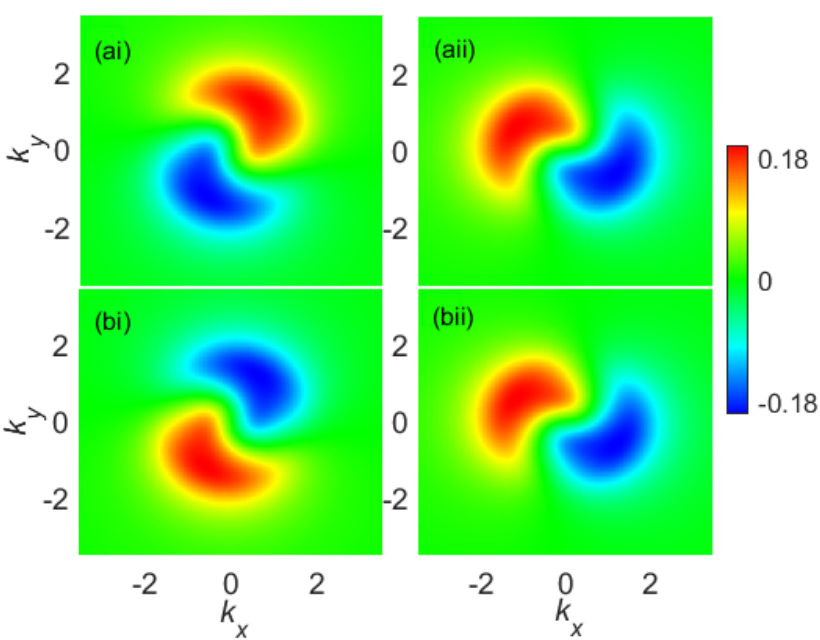} 
\caption{Plots of triplet pairing correlation (a)$_{L}\langle c_{-\textbf{k}\uparrow}c_{\textbf{k}\uparrow}\rangle_{R}$ and (b)$_{L}\langle c_{-\textbf{k}\downarrow}c_{\textbf{k}\downarrow}\rangle_{R}$, where (ai) and (bi) represent the real parts, (aii) and (bii) are the imaginary parts. The other parameters are $\alpha=1.5E_F/k_F$, $\Gamma=1E_F$ and $E_b=0.4E_F$}
\protect\label{fig:Kplane}
\end{figure}
 To investigate the influence of dissipation on the pairing signatures of the system, we calculated the singlet and triplet pairing correlations (see appendix~\ref{a:3} for details)
\begin{align}
F_s(\mathbf{k})&={}_L\!\langle
c_{-\mathbf{k}\downarrow}c_{\mathbf{k}\uparrow}\rangle_R,
&F_t^{\sigma\sigma}(\mathbf{k})&={}_L\!\langle
c_{-\mathbf{k}\sigma}c_{\mathbf{k}\sigma}\rangle_R.
\label{eq:pair_correlations}
\end{align}

Fig.~\ref{fig:Kplane}(a) and ~\ref{fig:Kplane}(b) show momentum-space contour plots of the triplet pairing fields. A particularly distinctive feature of Fig.~\ref{fig:Kplane} is the angular twisting of the triplet-pairing pattern in momentum space. Owing to rotational covariance, the triplet components can generally be written as
$F_t^{\sigma\sigma}(\mathbf{k})=
\mathcal{A}_{\sigma}(k)
e^{i s_{\sigma}\phi_{\mathbf{k}}}$, $
\mathcal{A}_{\sigma}(k)=
\left|\mathcal{A}_{\sigma}(k)\right|
e^{i\theta_{\sigma}(k)}$, $s_{\sigma}=\pm 1$,where $\phi_{\mathbf{k}}$ is the polar angle of $\mathbf{k}$, $\theta_{\sigma}(k)$  describes the phase, $s_{\sigma}$ specifies the chirality of the corresponding equal-spin
component. The phase of the triplet-pairing amplitude is therefore
$
\arg F_t^{\sigma\sigma}(\mathbf{k})
=
s_{\sigma}\phi_{\mathbf{k}}+\theta_{\sigma}(k)
$. In the Hermitian limit,$\theta_{\sigma}(k)=0,\pi$, the radial coefficient
$\mathcal{A}_{\sigma}(k)$  only depends on the magnitude of $\mathbf{k}$ and can be chosen to be real. Consequently, the angular orientation of the real and imaginary parts $F_t^{\sigma\sigma}(\mathbf{k})$ is
independent of the radial momentum, giving the familiar $p\pm ip$ texture~\cite{He2012}.

The situation changes qualitatively in the presence of dissipation. The right and left Bogoliubov coherence factors become intrinsically
complex, and the radial coefficient acquires a momentum-dependent phase. 
Hence, a radial variation of $\theta_{\sigma}(k)$ continuously rotates the
$p$-wave lobes as $k$ changes, producing the spiral-like angular texture
visible in Fig.~\ref{fig:Kplane}. In the 
$\uparrow-\uparrow$ channel [Figs.~\ref{fig:Kplane}(ai)-(aii)], the pairing field takes the form 
$p_y+\beta p_x+i(\beta p_y+p_x)$), while the 
$\downarrow-\downarrow$ channel [Figs.~\ref{fig:Kplane}(bi)-(bii)] is given by 
$-p_y-\beta p_x+i(\beta p_y+p_x)$
 ), where $\beta$ is a real number. The absolute pairing amplitudes of the two triplet pairings are therefore identical.
\begin{figure}[t] 
 \centering
\includegraphics[width=0.9\columnwidth]{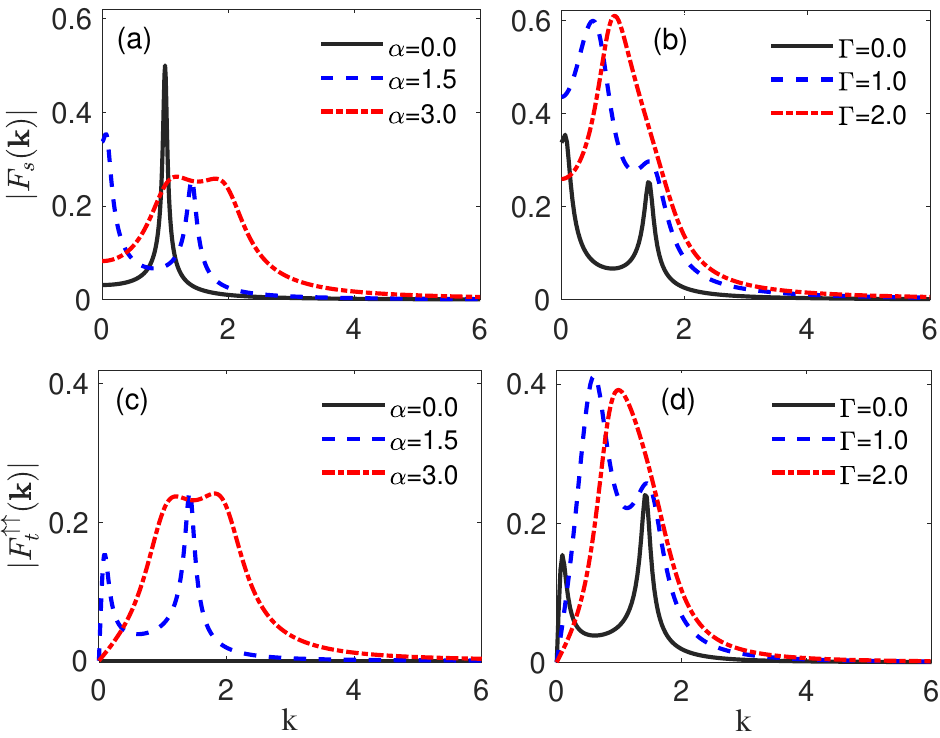} 
\caption{Plot of singlet pairing correlation (a-b) and triplet pairing correlation (c-d) for different Rashba strength $\alpha$ and dissipation $\Gamma$ at zero temperature. In both (a) and (c), $\Gamma=0$; (b) and (d), $\alpha=1.5$. $E_b=0.002E_F$}
\protect\label{fig:PairingCorrelations}
\end{figure}

Fig. \ref{fig:PairingCorrelations} shows the evolution of the singlet and triplet pairing correlations with $\alpha$ and $\Gamma$ in momentum space. In the absence of dissipation,  Rashba SOC generates finite triplet pairing components through spin-momentum locking; moreover, as SOC increases, the peak of the singlet and triplet pairing  spreads to wider momentum regime, as shown in Fig. \ref{fig:PairingCorrelations}(a) and \ref{fig:PairingCorrelations}(c). Interestingly, dissipation enhances the magnitude of both the singlet and triplet correlations, as shown in Fig. \ref{fig:PairingCorrelations}(b) and \ref{fig:PairingCorrelations}(d). This effect arises from the combined influence of SOC-induced spin mixing and dissipation, which alters the relative contribution of different helicity branches to the pairing wave function.

\begin{figure}[ht] 
 \centering
\includegraphics[width=0.9\columnwidth]{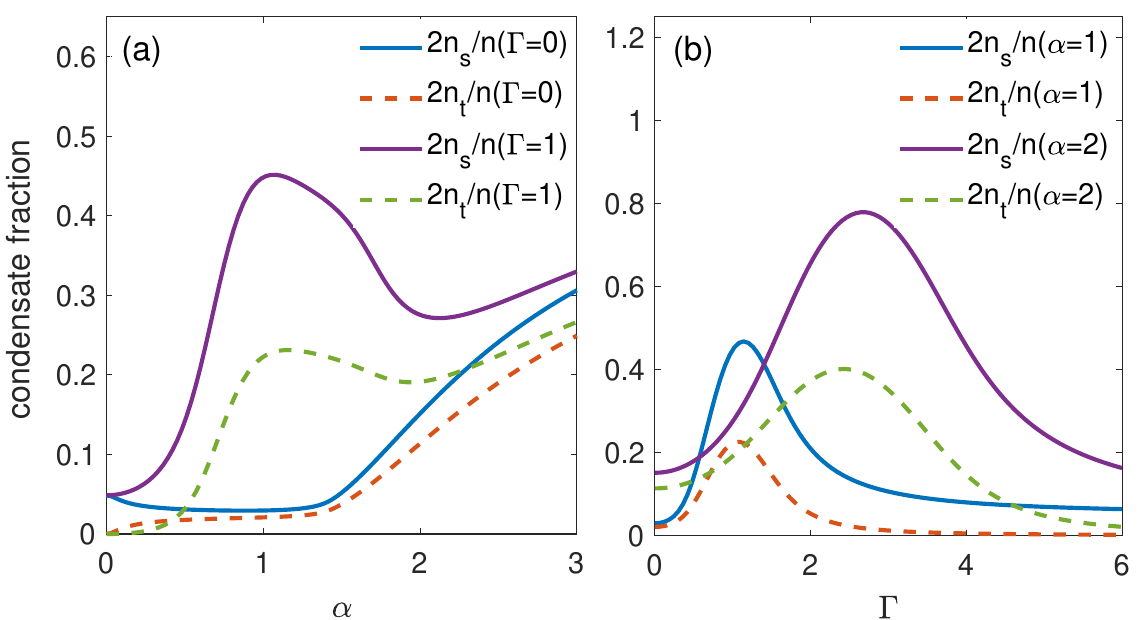} 
\caption{Plot of condensate fraction as a function of (a) $\alpha$ for different $\Gamma$ and (b) dissipation $\Gamma$ for different $\alpha$ at zero temperature. $E_b=0.002E_F$}
\protect\label{fig:np}
\end{figure}
 To  quantify the redistribution between pairing channels, we further calculate the condensate fractions of the singlet and triplet pairing contribution
$\frac{2n_s}{n}=\frac{2}{N}\sum_{\mathbf{k}}\left|F_s(\mathbf{k})\right|^2,
\frac{2n_t}{n}=\frac{2}{N}\sum_{\mathbf{k}}
|F_t^{\uparrow\uparrow}(\mathbf{k})|^2$~\cite{Anna2011,Shizhong2011}.  Fig.~\ref{fig:np}(a) shows the  singlet and triplet condensate fractions as functions of $\alpha$, which exhibit a similar behavior to that of the self-consistent pairing gap in Fig.~\ref{fig:Fig1Main}(a). Increasing dissipation enhances the triplet fraction before entering the strong-loss regime. This behavior demonstrates that dissipation can continuously tune the internal structure of Cooper pairs rather than merely changing the magnitude of the order parameter. The ability to manipulate singlet-triplet conversion through engineered dissipation provides a distinct advantage over conventional tuning methods. While interaction strength and SOC mainly control the energetic properties of the superfluid, spin-selective loss directly modifies the non-Hermitian quasiparticle wave functions and therefore offers a new pathway for designing unconventional pairing states.

\section{Conclusion}\label{conclusion} 
In this work, we investigated the NH mean-field ground state properties of a two-dimensional uniform $s$-wave Fermi superfluid subject to both Rashba SOC and spin-dependent atom loss. By self-consistently solving the NH gap and number equations, we uncovered a highly counterintuitive phenomenon:  moderate one-body dissipation can significantly enhance the superfluid order parameter $\Delta_0$, and by appropriately tuning the dissipation strength, one can continuously drive the BCS-BEC crossover in a spin-orbit coupled system. A particularly distinctive feature appears in the Bogoliubov quasi-particle spectrum. Depending on the self-consistent solutions, the exceptional condition can support either one or three exceptional ring superfluid regimes. Furthermore, we analyzed the momentum-space distribution of the pairing correlations and showed that the interplay between SOC and dissipation strongly modifies both the singlet and triplet pairing amplitudes. Our results demonstrate that spin-selective dissipation provides a versatile non-Hermitian control knob for manipulating superfluid pairing and crossover physics in spin-orbit coupling quantum gases. Such controllable non-Hermitian engineering may be accessible in ultracold atomic experiments using state-dependent optical pumping or Raman-assisted loss techniques. Our work therefore provides a general framework for exploring dissipative many-body phases in synthetic quantum systems.

\section*{Acknowledgments}
This work is supported by  the National Natural Science Foundation of China (Grant No. 12604414, 12374046 and 12634002), the Shanghai Science and Technology Innovation Action Plan (Grant No. 24LZ1400800), the State Key Laboratory of Micro-nano Engineering Science (Grant No. MES202605).
%

\appendix
\onecolumngrid
\renewcommand{\theequation}{A\arabic{equation}}
\setcounter{equation}{0}  

\section{Diagonalizing the Hamiltonian} \label{a:3}
In this section, we show the details of diagonalization and calculation of the pairing correlations. The mean-field Hamiltonian reads
\begin{align}
    H_{\text{MF}} &=\frac{1}{2}\sum_\textbf{k} \left(c^\dagger_{\textbf{k}\uparrow},c^\dagger_{\textbf{k}\downarrow},c_{-\textbf{k}\uparrow},c_{-\textbf{k}\downarrow}\right)
    \begin{pmatrix}
      \mathcal{H}_0(\textbf{k})&i\sigma_y\Delta_0 \\
        -i\sigma_y\Delta_0&-\mathcal{H}^\text{T}_0(-\textbf{k})   
    \end{pmatrix}
\left(c_{\textbf{k}\uparrow},c_{\textbf{k}\downarrow},c^\dagger_{-\textbf{k}\uparrow},c^\dagger_{-\textbf{k}\downarrow}\right)^\text{T}+\sum_{\textbf{k}}\xi_\textbf{k}-\frac{S\Delta_0^2}{g}\nonumber\\
&=\frac{1}{2}\sum_{\textbf{k}}\left[E^{+}_{\textbf{k}+}(\bar\alpha_{\textbf{k}\uparrow}\alpha_{\textbf{k}\uparrow}-\beta_{\textbf{k}\uparrow}\bar\beta_{\textbf{k}\uparrow})+E^{+}_{\textbf{k}-}(\bar\alpha_{\textbf{k}\downarrow}\alpha_{\textbf{k}\downarrow}-\beta_{\textbf{k}\downarrow}\bar\beta_{\textbf{k}\downarrow})\right]+\sum_{\textbf{k}}\xi_\textbf{k}-\frac{S\Delta_0^2}{g}\nonumber\\
\end{align}
where
\begin{align}
    E^\lambda_{\textbf{k}\pm}=\lambda\sqrt{\xi_\textbf{k}^2+\alpha^2k^2-\Gamma^2+\Delta_0^2\pm2\mathcal{B}_\textbf{k}}
\end{align}
is the eigenvalue of the BdG matrix in Eq.~\ref{eq2} and the Nambu spinor of the quasi-particle $(\alpha_{\textbf{k}\uparrow},\alpha_{\textbf{k}\downarrow},\bar\beta_{\textbf{k}\uparrow},\bar\beta_{\textbf{k}\downarrow})$ are related to $(c_{\textbf{k}\uparrow},c_{\textbf{k}\downarrow},c^\dagger_{-\textbf{k}\uparrow},c^\dagger_{-\textbf{k}\downarrow})$ with a similar transformation as
\begin{align}
c_{\mathbf{k}\uparrow}
&= u_{1\uparrow}\alpha_{\mathbf{k}\uparrow}
 + u_{2\uparrow}\alpha_{\mathbf{k}\downarrow}
 + u_{3\uparrow}\bar{\beta}_{\mathbf{k}\uparrow}
 + u_{4\uparrow}\bar{\beta}_{\mathbf{k}\downarrow}\nonumber\\
c_{\mathbf{k}\downarrow}
&= u_{1\downarrow}\alpha_{\mathbf{k}\uparrow}
 + u_{2\downarrow}\alpha_{\mathbf{k}\downarrow}
 + u_{3\downarrow}\bar{\beta}_{\mathbf{k}\uparrow}
 + u_{4\downarrow}\bar{\beta}_{\mathbf{k}\downarrow}\nonumber\\
c^\dagger_{-\mathbf{k}\uparrow}
&= v_{1\uparrow}\alpha_{\mathbf{k}\uparrow}
 + v_{2\uparrow}\alpha_{\mathbf{k}\downarrow}
 + v_{3\uparrow}\bar{\beta}_{\mathbf{k}\uparrow}
 + v_{4\uparrow}\bar{\beta}_{\mathbf{k}\downarrow}\nonumber\\
c^\dagger_{-\mathbf{k}\downarrow}
&= v_{1\downarrow}\alpha_{\mathbf{k}\uparrow}
 + v_{2\downarrow}\alpha_{\mathbf{k}\downarrow}
 + v_{3\downarrow}\bar{\beta}_{\mathbf{k}\uparrow}
 + v_{4\downarrow}\bar{\beta}_{\mathbf{k}\downarrow}
 \label{eqa3}
\end{align}
where these coefficients form the eigenvector $ |
        \Phi_{\mathbf{k}\eta}^{\lambda}
    \rangle_{R}$ of the BdG Hamiltonian.
\begin{align}
    M_{\mathbf{k}}
    |
        \Phi_{\mathbf{k}\eta}^{\lambda}
    \rangle_{R}
    &=
    E_{\mathbf{k}\eta}^{\lambda}
   |
        \Phi_{\mathbf{k}\eta}^{\lambda}
    \rangle_{R}
\end{align}
Similarly,
\begin{align}
c^\dagger_{\mathbf{k}\uparrow}
&= u'_{1\uparrow}\bar{\alpha}_{\mathbf{k}\uparrow}
 + u'_{2\uparrow}\bar{\alpha}_{\mathbf{k}\downarrow}
 + u'_{3\uparrow}\beta_{\mathbf{k}\uparrow}
 + u'_{4\uparrow}\beta_{\mathbf{k}\downarrow}\nonumber\\
c^\dagger_{\mathbf{k}\downarrow}
&= u'_{1\downarrow}\bar{\alpha}_{\mathbf{k}\uparrow}
 + u'_{2\downarrow}\bar{\alpha}_{\mathbf{k}\downarrow}
 + u'_{3\downarrow}\beta_{\mathbf{k}\uparrow}
 + u'_{4\downarrow}\beta_{\mathbf{k}\downarrow}\nonumber\\
c_{-\mathbf{k}\uparrow}
&= v'_{1\uparrow}\bar{\alpha}_{\mathbf{k}\uparrow}
 + v'_{2\uparrow}\bar{\alpha}_{\mathbf{k}\downarrow}
 + v'_{3\uparrow}{\beta}_{\mathbf{k}\uparrow}
 + v'_{4\uparrow}{\beta}_{\mathbf{k}\downarrow}\nonumber\\
c_{-\mathbf{k}\downarrow}
&= v'_{1\downarrow}\bar{\alpha}_{\mathbf{k}\uparrow}
 + v'_{2\downarrow}\bar{\alpha}_{\mathbf{k}\downarrow}
 + v'_{3\downarrow}{\beta}_{\mathbf{k}\uparrow}
 + v'_{4\downarrow}{\beta}_{\mathbf{k}\downarrow}
 \label{eqa5}
\end{align}
these coefficients form the corresponding left eigenvector $ |
        \Phi_{\mathbf{k}\eta}^{\lambda}
    \rangle_{L}$ and we have the biorthonormal normalization condition.
\begin{align}
     {}_L\langle
        \Phi_{\mathbf{k}\eta}^{\lambda}
        |
        \Phi_{\mathbf{k}\eta'}^{\lambda'}
    \rangle_{R}
    =
    \delta_{\eta\eta'}\delta_{\lambda\lambda'},
    \label{eq:app_biorthonormal}
\end{align}
Substituting Eq.~(\ref{eqa3}) and Eq.~(\ref{eqa5}) into Eq.~(\ref{eq:pair_correlations}), we obtain the expressions for the singlet and triplet pairing correlations.
\begin{align}
F_s(\mathbf{k})&=v'_{3\downarrow}u_{3\uparrow}+v'_{4\downarrow}u_{4\uparrow}\nonumber\\
F_t^{\uparrow\uparrow}(\mathbf{k})&=v'_{3\uparrow}u_{3\uparrow}+v'_{4\uparrow}u_{4\uparrow}\nonumber\\
F_t^{\downarrow\downarrow}(\mathbf{k})&=v'_{3\downarrow}u_{3\downarrow}+v'_{4\downarrow}u_{4\downarrow}\nonumber\\
\end{align}

\renewcommand{\theequation}{B\arabic{equation}}
\section{Numerical results in a strongly interacting regime \texorpdfstring{$E_b=0.9E_F$}{Eb = 0.9 Ef}}
\label{a:2}
In this section, we present the self-consistent numerical solutions for a more strongly interacting system with $E_b=0.9E_F$. As a benchmark, in the absence of both SOC and dissipation, the two-dimensional mean-field equations give
\begin{align}
    \Delta_0=\sqrt{2E_FE_b}\simeq 1.342E_F,
    \qquad
    \mu=E_F-\frac{E_b}{2}=0.55E_F .
\end{align}

\begin{figure}[h] 
 \centering
\includegraphics[width=0.48\columnwidth]{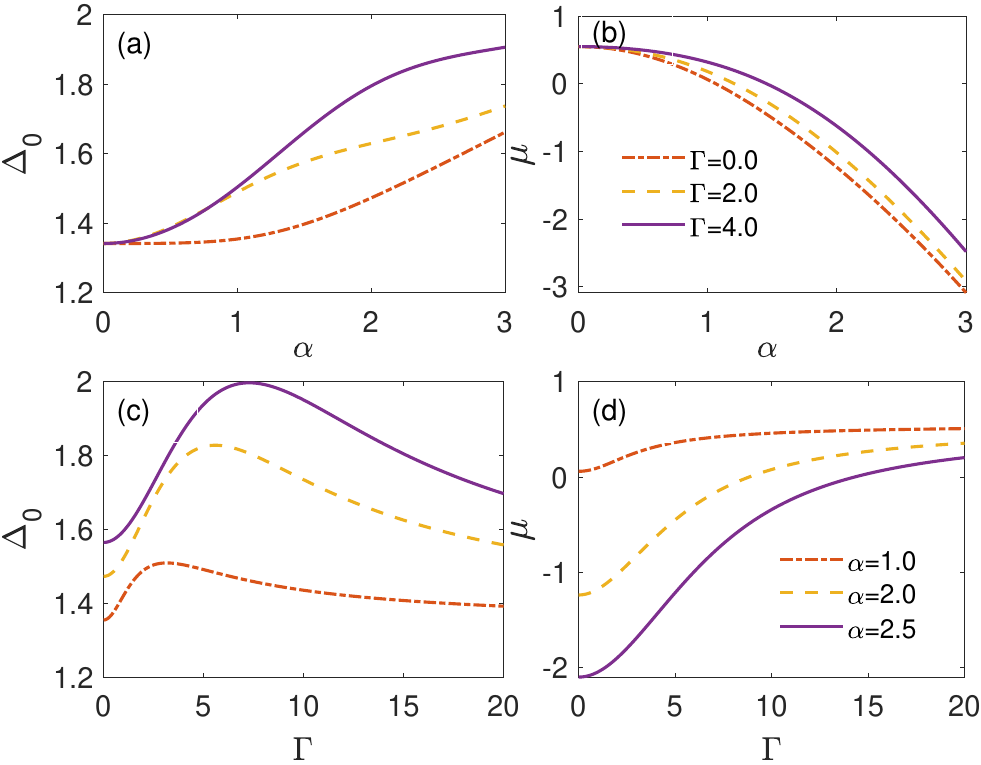} 
\caption{(a) Plot of $\Delta_0$ with respect to $\alpha$ for different $\Gamma$. (b) Plot of $\mu$ with respect to $\alpha$ for different $\Gamma$. (c) Plot of $\Delta$ with respect to $\Gamma$ for different $\alpha$. (d) Plot of $\mu$ with respect to $\Gamma$ for different $\alpha$. Here the interaction strength $E_b=0.9E_{F}$.}
\protect\label{fig:MainEb0.9}
\end{figure}
As shown in Fig.~8, the qualitative behavior found in the weakly interacting regime persists at stronger attraction: the pairing gap exhibits a nonmonotonic dependence on the dissipation strength, while the chemical potential is strongly modified by both SOC and dissipation. In particular, increasing $\Gamma$ initially enhances $\Delta_0$ but eventually suppresses this enhancement, whereas $\mu$ is driven upward toward its value in the absence of SOC and dissipation. These results indicate that the competition between Rashba SOC and spin-selective dissipation, as well as the resulting tunability of the superfluid state, is not restricted to the weak-coupling regime.


\begin{thebibliography}{25}%
\makeatletter
\providecommand \@ifxundefined [1]{%
 \@ifx{#1\undefined}
}%
\providecommand \@ifnum [1]{%
 \ifnum #1\expandafter \@firstoftwo
 \else \expandafter \@secondoftwo
 \fi
}%
\providecommand \@ifx [1]{%
 \ifx #1\expandafter \@firstoftwo
 \else \expandafter \@secondoftwo
 \fi
}%
\providecommand \natexlab [1]{#1}%
\providecommand \enquote  [1]{``#1''}%
\providecommand \bibnamefont  [1]{#1}%
\providecommand \bibfnamefont [1]{#1}%
\providecommand \citenamefont [1]{#1}%
\providecommand \@href[1]{\@@startlink{#1}\@@href}%
\providecommand \@@href[1]{\endgroup#1\@@endlink}%
\providecommand \@sanitize@url [0]{\catcode `\\12\catcode `\$12\catcode
  `\&12\catcode `\#12\catcode `\^12\catcode `\_12\catcode `\%12\relax}%
\providecommand \@@startlink[1]{}%
\providecommand \@@endlink[0]{}%
\providecommand \@url [1]{\endgroup\@href {#1}{\urlprefix }}%
\providecommand \urlprefix  [0]{URL }%
\providecommand \doibase [0]{https://doi.org/}%
\providecommand \selectlanguage [0]{\@gobble}%
\providecommand \bibinfo  [0]{\@secondoftwo}%
\providecommand \bibfield  [0]{\@secondoftwo}%
\providecommand \translation [1]{[#1]}%
\providecommand \BibitemOpen [0]{}%
\providecommand \bibitemStop [0]{}%
\providecommand \bibitemNoStop [0]{.\EOS\space}%
\providecommand \EOS [0]{\spacefactor3000\relax}%
\providecommand \BibitemShut  [1]{\csname bibitem#1\endcsname}%
\let\auto@bib@innerbib\@empty
\bibitem{Ashida2020}  Y. Ashida, Z. Gong and M. Ueda, \textit{Advances in Physics} {\bf 69(3)}, 249-435 (2020).

\bibitem{Ganainy2018}  R. El-Ganainy, K. Makris, M. Khajavikhan, et al., \textit{Nat. Phys.} {\bf 14}, 11-19 (2018).

\bibitem{Ueda2019}  K. Yamamoto,  M. Nakagawa,  K. Adachi,  K. Takasan,  M. Ueda, N. Kawakami, \textit{Phys. Rev. Lett.} {\bf 123}, 123601 (2019).

\bibitem{Iskin2021}  M. Iskin, \textit{Phys. Rev. A.} {\bf 103}, 013724 (2021).

\bibitem{Shi2024}  T. Shi, S. Wang, Z. Zheng, W. Zhang, \textit{Phys. Rev. A.} {\bf 109}, 063306 (2024).

\bibitem{koga2024}  S. Takemori, K. Yamamoto, A. Koga, \textit{Phys. Rev. B.} {\bf 109}, L060501 (2024).

\bibitem{Wang2025}  S. Wang, Y. Shi and W. Zhang, \textit{Front. Phys.} {\bf 21(6)}, 065201 (2025).

\bibitem{zhu2026}  P. Zhu, L. Zhou, J. Zhong, \textit{iScience} {\bf 29}, 116686 (2026).

\bibitem{Okugawa2019}  R. Okugawa and T. Yokoyama,  \textit{Phys. Rev. B.} {\bf 99}, 041202 (2019).

\bibitem{Kozii2024}  V. Kozii, L. Fu, \textit{Phys. Rev. B.} {\bf 109}, 235139 (2024).

\bibitem{wang2018} S. Yao, Z. Wang, \textit{Phys. Rev. Lett.} {\bf 121}, 086803 (2018).

\bibitem{Lee2019}  C. H. Lee, and R. Thomale, \textit{Phys. Rev. B.} {\bf 99}, 201103(R) (2019).

\bibitem{Gon2020}  L. Li, C. H. Lee, S. Mu, J. Gong, \textit{Nat. Commun.} {\bf 11}, 5491 (2020).

\bibitem{Wang2026}  S. Wang, W. Xiong, Z. Zhang, Y. Cheng, X. Liu, \textit{Phys. Rev. Lett.} {\bf 136}, 026601 (2026).

\bibitem{p2020}  X. M. Yang and Z. Song, \textit{Phys. Rev. A.} {\bf 102}, 022219 (2020).
 
\bibitem{Ji2025}  J. Ji, W. Nie, \textit{arXiv:} {\bf 2508}, 12360 (2025).

\bibitem{Gong2011}  M. Gong, S. Tewari and C. Zhang, \textit{Phys. Rev. Lett.} {\bf 107}, 195303 (2011).

\bibitem{Jiang2011}  L. Jiang, X. liu, H. Hu and H. Pu, \textit{Phys. Rev. A.} {\bf 84}, 063618 (2011).

\bibitem{Ren2022}  Z. Ren, D. Liu E. Zhao, et al. \textit{Nat. Phys.} {\bf 18}, 385 (2022).

\bibitem{Hu2011}  H. Hu, L. Jiang, X. Liu and H. Pu, 
\textit{Phys. Rev. Lett.} {\bf 107}, 195304 (2011).

\bibitem{Zhai2011} Z. Yu and H. Zhai, \textit{Phys. Rev. Lett.} {\bf 107}, 195305 (2011).

\bibitem{He2012}  L. He and X. Huang, \textit{Phys. Rev. Lett.} {\bf 108}, 145302 (2012).

\bibitem{Shi2016}  H. Shi, P. Rosenberg, S. Chiesa and S. Zhang, \textit{Phys. Rev. Lett.} {\bf 117}, 040401 (2016).

\bibitem{Iskin2026} R. N. Kalkan, and M. Iskin, \textit{arxiv:} {\bf 2608}, 21110 (2026).

\bibitem{Zhou2019} L. Zhou and X. Cui, \textit{iScience} {\bf 14}, 257 (2019).

\bibitem{Zhao2023}  X. Zhao and L. Zhou, \textit{Phys. Rev. A.} {\bf 108}, 013311 (2023).

\bibitem{Zhou2020}  L. Zhou, W. Yi, X. Cui, \textit{Phys. Rev. A.} {\bf 102}, 043310 (2020).

\bibitem{Zhou2022}  L. Zhou, H. Li, W. Yi and X. Cui, \textit{Commun. Phys.} {\bf 5}, 252 (2022).

\bibitem{Li2023}  H. Li,  W. Yi, \textit{Phys. Rev. A.} {\bf 107}, 013306 (2023).

\bibitem{Li2024}  Z. Lei, C. H. Lee, L. Li, \textit{Commun. Phys.} {\bf 7}, 100 (2024).

\bibitem{nat2025} E. Zhao, Z. Wang, C. He, T. F. J. Poon, K. K. Pak, Y.-J. Liu, P. Ren, X.-J. Liu, G.-B. Jo, \textit{Nature} {\bf 637}, 565-573 (2025).

\bibitem{Takemori2024}  S. Takemori, K. Yamamoto and A. Koga, \textit{Phys. Rev. B.} {\bf 110}, 184518 (2024).

\bibitem{Brody2014}  D. Brody, \textit{J. Phys. A:  Math. Theor.} {\bf 47}, 035305 (2014).

\bibitem{Takemori2025} S. Takemori, K. Yamamoto and A. Koga, \textit{Phys. Rev. Lett.} {\bf 135}, 266002 (2025).

\bibitem{Shizhong2011}  J. Vyasanakere, S. Zhang and V. Shenoy, \textit{Phys. Rev. B.} {\bf 84}, 014512 (2011).

\bibitem{Anna2011} L. Dell'Anna, G. Mazzarella, L. Salasnich, \textit{Phys. Rev. A.} {\bf 84}, 033633 (2011).








\end{thebibliography}
\end{document}